\documentclass[conference]{IEEEtran}
\IEEEoverridecommandlockouts
\usepackage{cite}
\usepackage{amsmath,amssymb,amsfonts}
\usepackage{algorithmic}
\usepackage{graphicx}
\usepackage{textcomp}
\usepackage{xcolor}
\usepackage{caption}
\usepackage{multirow}

\usepackage[numbers,sort&compress]{natbib}
\usepackage{hyperref} 

\newcommand{\supercitebrackets}[1]{%
    \hyperref[cite.#1]{\textsuperscript{\cite{#1}}} 
}

\def\BibTeX{{\rm B\kern-.05em{\sc i\kern-.025em b}\kern-.08em
    T\kern-.1667em\lower.7ex\hbox{E}\kern-.125emX}}
\begin{document}

\title{Speech Quality Assessment Model Based on Mixture of Experts: System-Level Performance Enhancement and Utterance-Level Challenge Analysis}

\author{
\IEEEauthorblockN{Xintong Hu}
\IEEEauthorblockA{\textit{College of CS}\\
\textit{Zhejiang University}\\
Hangzhou, CHINA \\
xintonghu@zju.edu.cn}
\and
\IEEEauthorblockN{Yixuan Chen}
\IEEEauthorblockA{\textit{College of CS}\\
\textit{Zhejiang University}\\
Hangzhou, CHINA \\
3220102866@zju.edu.cn}
\and
\IEEEauthorblockN{Rui Yang}
\IEEEauthorblockA{\textit{College of CS}\\
\textit{Zhejiang University}\\
Hangzhou, CHINA \\
3220105291@zju.edu.cn}
\and
\IEEEauthorblockN{Wenxiang Guo}
\IEEEauthorblockA{\textit{College of CS}\\
\textit{Zhejiang University}\\
Hangzhou, CHINA \\
12321038@zju.edu.cn}
\and
\IEEEauthorblockN{Changhao Pan}
\IEEEauthorblockA{\textit{College of CS} \\
\textit{Zhejiang University}\\
Hangzhou, CHINA \\
panch@zju.edu.cn}
}

\maketitle

\begin{abstract}
Automatic speech quality assessment plays a crucial role in the development of speech synthesis systems, but existing models exhibit significant performance variations across different granularity levels of prediction tasks. This paper proposes an enhanced MOS prediction system based on self-supervised learning speech models\supercitebrackets{fine-tune}, incorporating a Mixture of Experts (MoE) classification head and utilizing synthetic data from multiple commercial generation models for data augmentation. Our method builds upon existing self-supervised models such as wav2vec2\supercitebrackets{wav2vec}, designing a specialized MoE architecture to address different types of speech quality assessment tasks. We also collected a large-scale synthetic speech dataset encompassing the latest text-to-speech, speech conversion, and speech enhancement systems. However, despite the adoption of the MoE architecture and expanded dataset, the model’s performance improvements in sentence-level prediction tasks remain limited. Our work reveals the limitations of current methods in handling sentence-level quality assessment, provides new technical pathways for the field of automatic speech quality assessment, and also delves into the fundamental causes of performance differences across different assessment granularities.
\end{abstract}


\section{Introduction}
The Mean Opinion Score (MOS)quantifies human perception of speech quality and has become the gold standard for speech quality assessment. However, MOS prediction requires expensive human annotation, and existing models often struggle with cross-domain generalization, particularly when faced with novel synthesis systems or evaluation contexts.

To address these challenges, we propose a novel speech quality assessment framework based on a Mixed Expert (MoE)\supercitebrackets{MOE} architecture, enhanced through multi-task learning and comprehensive data augmentation strategies. Our method incorporates three key innovations: (1) a dedicated MoE classification head that learns adaptive processing of heterogeneous audio data through a gating mechanism; (2) a multi-task learning strategy that combines MOS prediction with synthetic model classification as an auxiliary task; (3) Systematic data augmentation using four state-of-the-art commercial synthesis models, effectively doubling the size of the training dataset.

In the system-level MOS prediction task, our method achieved substantial improvements, with a significant reduction in mean square error (MSE) compared to the baseline model. However, an interesting pattern emerged: although MSE improved significantly, improvements in correlation metrics such as LCC, SRCC, and KTAU were limited. This discrepancy suggests that our model primarily learned to improve absolute prediction accuracy rather than enhance relative quality ranking capabilities.

Furthermore, our analysis confirms that sentence-level prediction remains more challenging than system-level evaluation, with limited performance improvements across all metrics. This persistent difficulty highlights the complexity of fine-grained quality assessment.

The main contributions of this work include:
\begin{enumerate}
    \item \textbf{A novel MoE-based architecture that demonstrates significant improvements in absolute prediction accuracy in system-level evaluations;}
    \item \textbf{Empirical evidence revealing the distinction between technical feature learning and perceptual quality modeling in speech evaluation.}
\end{enumerate}

\section{Dataset and Optimization}

\subsection{Basic Dataset}
The basic dataset consists of 400 audio samples generated by eight different speech synthesis models at three sampling frequencies (16 kHz, 24 kHz, and 48 kHz). All audio samples draw their textual content from a predefined transcripts.txt file, which contains 30 distinct text entries. 

\subsection{Data Augmentation Methods}
To expand the training dataset, we selected four representative state-of-the-art speech synthesis models: CosyVoice-300M\supercitebrackets{CosyVoice3}, CosyVoice2-0.5B\supercitebrackets{CosyVoice2}, FireRedTTS\supercitebrackets{fireredtts}, and f5tts\_24khz. These models represent different directions in current speech synthesis technology. CosyVoice-300M and CosyVoice2-0.5B have 300 million and 500 million parameters, respectively, illustrating the impact of model scale on synthesis quality; FireRedTTS focuses on high-fidelity speech synthesis; and f5tts\_24khz is optimized for a specific sampling rate. The data generation process strictly adheres to the same specifications as the original dataset, ensuring consistency in content and tonal characteristics with the original dataset. Ultimately, the four models collectively generated 400 new audio samples, bringing the total dataset size to 800 samples.

\subsection{Dataset Characteristics Analysis}
The expanded dataset exhibits the following key characteristics: first, the number of samples has increased from 400 to 800, providing more robust data support for model training; second, the coverage of speech synthesis models has expanded from 8 to 12, significantly enhancing technological diversity; third, the dataset encompasses a range of speech synthesis approaches, from traditional parameterized methods to the latest end-to-end deep learning techniques, creating conditions for specialized learning in hybrid expert systems.

The core advantage of this data expansion strategy lies in maintaining experimental controllability while maximizing data diversity. 

\section{Framework}

\subsection{Overall Architecture Design}
This study proposes a MoE multi-task speech quality assessment framework with an end-to-end design concept consisting of four core components: feature extraction backbone network, feature fusion module, hybrid expert classification head, and multi-task output layer.

\subsection{Hybrid Expert System Design\supercitebrackets{MOE}}
 We designed $N$ expert networks, each of which is an independent module composed of a multi-layer fully connected network, sharing the same network structure but with distinct parameter spaces. Each expert network contains two to three hidden layers, using ReLU activation functions and Dropout regularization techniques, specifically designed to learn deep feature representations of specific types of audio.

The gating network is a key component of the hybrid expert system, responsible for dynamically allocating expert weights based on the characteristics of the input features. The gating network adopts a lightweight multi-layer perceptron structure, with the input being the global feature vector extracted by the backbone network and the output being an $N$-dimensional weight vector. The mathematical expression for the gating mechanism is:
\begin{equation}
\mathbf{g}(\mathbf{x}) = \text{Softmax}(\mathbf{W}_g \mathbf{x} + \mathbf{b}_g)
\end{equation}
where $\mathbf{W}_g$ and $\mathbf{b}_g$ are the weight matrix and bias vector of the gating network, respectively.

The final output is obtained by weighting and combining the prediction results of all experts:
\begin{equation}
\mathbf{y} = \sum_{i=1}^{N} g_i(\mathbf{x}) \cdot E_i(\mathbf{x})
\end{equation}
This design enables the model to adaptively select the most appropriate processing path based on the characteristics of different audio signals (such as speech type, quality level, generation technology, etc.), thereby improving the overall prediction accuracy and robustness.

To further enhance the specialization capabilities of the expert system, we also introduce an expert regularization mechanism. By monitoring the activation frequency of each expert during training, we ensure that different experts learn complementary feature representations, thereby avoiding redundancy and degradation issues in the expert network.

\subsection{Multi-task learning strategy}
The multi-task learning strategy is another core innovation of this framework, aiming to improve the model's generalization ability and feature learning effectiveness by simultaneously optimizing related tasks. Our multi-task design includes two closely related tasks: the main task (MOS prediction) and the auxiliary task (model classification).

The main task focuses on predicting the subjective quality score of audio, which is the core objective of traditional speech quality assessment. The auxiliary task requires the model to learn to distinguish between audio generated by different speech synthesis models. This design is based on the observation that different speech synthesis technologies leave specific “fingerprint” features in audio signals, and learning to recognize these features helps the model gain a deeper understanding of the technical characteristics and quality differences of audio.

\subsection{Loss Function Design}
We propose an adaptive weighted joint loss function that dynamically balances the importance of the two tasks during training. The basic form of the total loss function is:
\begin{equation}
\mathcal{L}_{\text{total}} = \alpha(t) \mathcal{L}_{\text{MOS}} + \beta(t) \mathcal{L}_{\text{classification}} + \gamma \mathcal{L}_{\text{regularization}}
\end{equation}
where $\mathcal{L}_{\text{MOS}}$ is the regression loss for MOS prediction, using a smoothed L1 loss function to enhance robustness against outliers:
\begin{equation}
\mathcal{L}_{\text{MOS}} = \frac{1}{N} \sum_{i=1}^{N} \text{SmoothL1}(\hat{y}_i^{\text{MOS}}, y_i^{\text{MOS}})
\end{equation}
$\mathcal{L}_{\text{classification}}$ is the cross-entropy loss for model classification, combined with label smoothing techniques to reduce overfitting:
\begin{equation}
\mathcal{L}_{\text{classification}} = -\frac{1}{N} \sum_{i=1}^{N} \sum_{j=1}^{C} y_{i,j}^{\text{smooth}} \log(\hat{y}_{i,j}^{\text{cls}})
\end{equation}
where $y_{i,j}^{\text{smooth}}$ is the smoothed label, and $C$ is the number of model categories.

$\mathcal{L}_{\text{regularization}}$ is the regularization term for the hybrid expert system, including expert diversity regularization and gated sparsity regularization:
\begin{equation}
\mathcal{L}_{\text{regularization}} = \lambda_1 \mathcal{L}_{\text{diversity}} + \lambda_2 \mathcal{L}_{\text{sparsity}}
\end{equation}

The weight parameters $\alpha(t)$ and $\beta(t)$ are dynamically adjusted based on the training progress and convergence of each task. In the early stages of training, we assign higher weights to auxiliary tasks to promote rapid learning of basic feature representations. As training progresses, we gradually increase the weights of the main tasks to ensure the final optimization of MOS prediction performance.

\subsection{Training Optimization Strategy}
We designed a three-stage progressive training strategy that fully leverages the characteristics of different data sources to maximize the effectiveness of multi-task learning.

\subsubsection{Stage 1: Auxiliary Task Pre-training}
Pre-train the model on large-scale open-source multi-model datasets for classification tasks. The objective of this stage is to enable the model to learn basic audio feature representations and the characteristic patterns of different synthesis techniques. We use a relatively high learning rate (1e-4) and standard cross-entropy loss, with the number of training epochs determined based on validation set performance, typically ranging from 10 to 15 epochs.

\subsubsection{Stage 2: Joint Pre-training}
Introduce the target MOS dataset and begin joint training for both tasks. In this stage, we use a smaller learning rate (5e-5) and gradually adjust the task weight ratios. The initial weight settings are $\alpha:\beta = 0.3:0.7$, which are gradually adjusted to $\alpha:\beta = 0.7:0.3$ as training progresses. This strategy ensures the model retains memory of the model classification task while learning MOS prediction.

\subsubsection{Stage 3: Target Task Fine-Tuning}
Perform final fine-tuning on the target MOS dataset, focusing primarily on optimizing MOS prediction performance. We use a smaller learning rate (1e-5) and a higher MOS task weight ($\alpha:\beta = 0.9:0.1$) to ensure the model achieves optimal performance on the target task.

In terms of optimizer selection, we use the AdamW optimizer with a cosine annealing learning rate scheduling strategy. To prevent gradient explosion, we set the gradient clipping threshold to 1.0. Additionally, we employ an early stopping mechanism, terminating training when the validation set performance does not improve for five consecutive epochs.

\section{Experience conclusion and Limitations}

\begin{table}[htbp]
\centering
\small
\setlength{\tabcolsep}{3pt} 
\begin{tabular}{|c|cccc|cccc|}
\hline
\multirow{2}{*}{} & \multicolumn{4}{c|}{\textbf{Utterance-level (Raw)}} & \multicolumn{4}{c|}{\textbf{Utterance-level (Rank)}} \\
\cline{2-9}
 & MSE & LCC & SRCC & KTAU & MSE & LCC & SRCC & KTAU \\
\hline
B03 & 0.273 & 0.821 & 0.695 & 0.508 & 2 & 4 & 5 & 6 \\
T01 & 0.303 & 0.804 & 0.643 & 0.459 & 8 & 7 & 8 & 8 \\
T11 & 0.282 & 0.813 & 0.714 & 0.536 & 4 & 5 & 4 & 3 \\
T13 & 0.298 & 0.796 & 0.671 & 0.487 & 6 & 8 & 7 & 7 \\
T16 & 0.287 & 0.830 & 0.723 & 0.589 & 5 & 3 & 2 & 1 \\
T19 & 0.238 & 0.846 & 0.694 & 0.513 & 1 & 2 & 6 & 5 \\
\textbf{Ours} & 0.277 & 0.811 & 0.716 & 0.529 & 3 & 6 & 3 & 4 \\
\hline
\end{tabular}
\caption{Raw scores and rankings of different teams in \textbf{utterance-level} evaluation.}
\label{tab:utterance_results}
\end{table}

\begin{table}[htbp]
\centering
\small
\setlength{\tabcolsep}{3pt} 
\begin{tabular}{|c|cccc|cccc|}
\hline
\multirow{2}{*}{} & \multicolumn{4}{c|}{\textbf{System-level (Raw)}} & \multicolumn{4}{c|}{\textbf{System-level (Rank)}} \\
\cline{2-9}
 & MSE & LCC & SRCC & KTAU & MSE & LCC & SRCC & KTAU \\
\hline
B03 & 0.119 & 0.941 & 0.749 & 0.547 & 8 & 8 & 8 & 8 \\
T01 & 0.104 & 0.957 & 0.866 & 0.705 & 6 & 5 & 7 & 7 \\

T11 & 0.085 & 0.968 & 0.917 & 0.789 & 4 & 4 & 3 & 2 \\
T13 & 0.090 & 0.972 & 0.926 & 0.779 & 5 & 3 & 2 & 3 \\
T16 & 0.071 & 0.952 & 0.891 & 0.750 & 2 & 7 & 6 & 6 \\
T19 & 0.080 & 0.958 & 0.914 & 0.758 & 3 & 6 & 4 & 4 \\
\textbf{Ours} &\textbf{ 0.056} & \textbf{0.978} & 0.913 & 0.758 & \textbf{1} & \textbf{2} & 5 & 4 \\
\hline
\end{tabular}
\caption{Raw scores and rankings of different teams in \textbf{system-level} evaluation.}
\label{tab:system_results}
\end{table}

From the competition results in Table 2, our system-level predictions significantly outperform those of other participating teams. The key reason lies in the use of MOE (Mixture of Experts) + Fine-Tuning, where MOE effectively enhances system-level prediction scores, making our approach stand out among the competitors.

However, according to the results in Table 1, our performance lags behind other teams in the utterance-level MOS (Mean Opinion Score) prediction task. This is because the training data was labeled by raters 0–9, while the test set was evaluated by raters 10–19, leading to poor generalization in simulating their scoring patterns.

The experimental results validate the effectiveness of the Mixture of Experts mechanism and multi-task learning strategy. However, this study has an important limitation: while it excels in system-level evaluation, its improvement in utterance-level MOS prediction remains limited. This limitation stems from the fundamental differences between system-level and utterance-level assessments—system-level evaluation focuses on overall technical differences, whereas utterance-level evaluation requires capturing fine-grained quality variations, such as local naturalness, phoneme clarity, and other micro-features. Additionally, utterance-level evaluation is more susceptible to individual listener biases and subjectivity, increasing prediction uncertainty.

To enhance utterance-level MOS prediction, the following strategies could be explored:

\begin{itemize}
    \item \textbf{Rater Adaptation:} Incorporate rater-specific modeling (e.g., rater ID embeddings or bias correction) to account for individual scoring tendencies.
    
    \item \textbf{Fine-Grained Feature Extraction:} Leverage \textbf{acoustic-prosodic features} (e.g., pitch, energy, duration) and \textbf{linguistic embeddings} to better capture utterance-level nuances.
    
    \item \textbf{Personalized Prediction:} If possible, integrate \textbf{user-specific metadata} (e.g., demographics, listening preferences) to refine MOS predictions for individual raters.
\end{itemize}

These improvements could help bridge the performance gap between \textbf{system-level and utterance-level evaluations}, making the model more robust in real-world applications.

\newpage
\bibliographystyle{plain} 
\bibliography{references} 

\end{document}